\documentclass[10pt,fleqn]{article}
\usepackage{amssymb}

\newcommand{\name}[1]{\begin{flushleft}
                       \LARGE \bf #1
                       \end{flushleft}\vspace{-3mm}}

\newcommand{\Author}[1]{\begin{flushleft}
                       \it #1 \end{flushleft}}

\newcommand{\Adress}[1]{\begin{flushleft}
                       \it #1 \end{flushleft}}

\newcommand{\be}{\begin{equation}}
\newcommand{\ee}{\end{equation}}
\newcommand{\ba}{\hspace*{-5pt}\begin{array}}
\newcommand{\ea}{\end{array}}
\newcommand{\p}{\partial}
\newcommand{\ds}{\displaystyle}

\newcommand{\pbf}[1]{\mbox{\mathversion{bold}$#1$}}

\begin{document}

\name{On the possible types of equations\\
 for zero-mass particles}

\medskip

\noindent{published in {\it Lettere al Nuovo Cimento},  1973,  {\bf 7}, N 11, P. 439--442.}

\Author{Wilhelm I. FUSHCHYCH and Anatoly G. NIKITIN}

\Adress{Institute of Mathematics of the National Academy of
Sciences of Ukraine, \\ 3 Tereshchenkivska  Street, 01601 Kyiv-4,
UKRAINE}

\noindent {\tt http://www.imath.kiev.ua/\~{}appmath/wif.html\\
http://www.imath.kiev.ua/\~{}nikitin\\ E-mails:
symmetry@imath.kiev.ua,  nikitin@imath.kiev.ua}

\bigskip

\noindent
A number of papers dedicated to the description of free particles and antiparticles
with zero mass and spin $\frac 12$ has recently appeared [1--6].

A great many equations with different $C$, $P$, $T$ properties have been proposed and
the impression could be formed that there are many nonequivalent theories for
zero-mass particles. The purpose or this paper is to show that it is not the case and
to describe all nonequivalent equations.

{\bf 1.} First we shall formulate the result [1] obtained for a particle of spin $\frac 12$
in such a form that all principal assertions will be valid for massless particles
of arbitraly spin. It has been shown~[1] that for a particle of spin $\frac 12$
three types of nonequivalent two-component Poincar\'e-invariant equations exist.
These three of equations are equivalent to the Dirac equation
\be
i\frac{\p \Psi(t,\pbf{x})}{\p t} ={\mathcal H} \Psi(t,\pbf{x}),
\qquad {\mathcal H}=\gamma_0 \gamma_a p_a, \quad a=1,2,3,
\ee
with one out of three (actually, one out of six) subsidiary conditions imposed on a wave
function
\be
P_1^+\Psi=0 \qquad \mbox{or} \qquad P_1^-\Psi=0, \qquad
P_1^\pm =\frac 12 (1\pm i\gamma_4), \qquad \gamma_4=-\gamma_0\gamma_1\gamma_2
\gamma_3,
\ee
\be
P_2^+\Psi=0 \qquad \mbox{or} \qquad P_2^-\Psi=0, \qquad
P_2^\pm =\frac 12 (1\pm i\gamma_4\hat \varepsilon), \qquad \hat \varepsilon =
\frac{\mathcal H}{E},
\ee
\be
P_3^+\Psi=0 \qquad \mbox{or} \qquad P_3^-\Psi=0, \qquad
P_3^\pm =\frac 12 (1\pm \hat \varepsilon), \qquad E =
\sqrt{p_1^2+p_2^2 +p_2^2}.
\ee
Conditions (2)--(4) are Poincar\'e invariant since the projection operators
$P_a^\pm$ commute with the generators of the Poincar\'e group $P(1,3)$
\be
\ba{l}
\ds P_0={\mathcal H}=\gamma_0 \gamma_ap_a, \qquad P_a=p_a=-i\frac{\p}{\p x_a},
\qquad\! J_{0a}=tp_a -\frac 12 (x_a P_0+P_0 x_a),
\vspace{2mm}\\
\ds J_{ab}=x_a p_b -x_b p_a +S_{ab}, \qquad
S_{ab}=\frac i4 (\gamma_a \gamma_b -\gamma_b \gamma_a).
\ea\!
\ee
It should be emphasized that only the operator $P_1^\pm$
is local in co-ordinate space. If we introduce the four-component (as a matter
of fact, two-component) modes
\be
\chi_a^\pm =P_a^\pm \Psi,
\ee
equations (1) with subsidiary conditions (2)--(4) can be written in the form
\be
i\frac{\p \chi_a^\pm}{\p t} =(\gamma_0\gamma_b p_b \pm \varkappa_a \gamma_0 P_a^\mp)
\chi_a^\pm,
\ee
where $\varkappa_a$  are arbitrary constants. The wave functions
$\chi_a^\pm$ satisfy conditions (2)--(4) automatically. One of the equations (7), namely
the equation for $\chi_1^+$ (or $\chi_1^-$),
is equivalent, as is well known, to the two-component Weyl equation. Subsidiary
conditions (2)--(4) have been generalized in~[7] to massless particles of arbitrary spin
starting from the $2(2s+1)$-component equation.

These results are almost evident from the group-theoretical point of view. Indeed, on the set
$\{\Psi\}$ of solutions of the equation (1) the following direct sum of irreducible
representations of the group $P(1,3)$ is realized:
\be
D^+(\lambda=1)\oplus D^-(\lambda=-1)\oplus D^+(\lambda=-1)\oplus D^-(\lambda=1),
\ee
where $D^\varepsilon(\lambda)$ is the one-dimensional irreducible representation of the
$P(1,3)$ group characterized by the eigenvalue $\varepsilon=\pm 1$
of the sign energy operator $\hat \varepsilon$ and by the eigenvalue $\lambda=\pm 1$
of the helicity operator
\be
\hat \Lambda=2\frac{J_{12} P_3 +J_{23} P_1 +J_{01} P_2}{E}=i\gamma_4 \hat\varepsilon.
\ee
Two-dimensional subspaces of representations
\be
D^+(\lambda=1)\oplus D^-(\lambda=-1) \qquad \mbox{or} \qquad
D^+(\lambda=-1)\oplus D^-(\lambda=1),
\ee
\be
D^+(\lambda=1)\oplus D^-(\lambda=1) \qquad \mbox{or} \qquad
D^+(\lambda=-1)\oplus D^-(\lambda=-1),
\ee
\be
D^+(\lambda=1)\oplus D^+(\lambda=-1) \qquad \mbox{or} \qquad
D^-(\lambda=1)\oplus D^-(\lambda=-1),
\ee
are selected by subsidiary conditions (2)--(4) from $\{\Psi\}$ in a Poincar\'e-invariant manner.

The operators $P$, $T$, $C$ (their definitions see e.g. in [8])
and $\hat \Lambda$, $\hat \varepsilon$ satisfy the relations
\be
[P^{(1)}, \hat \Lambda]_+=[P^{(1)},\hat \varepsilon]_-=[T^{(2)},\hat \Lambda]_-=
[T^{(2)},\hat\varepsilon]_-=[C,\hat\Lambda]_-=[C,\hat\varepsilon]_+=0.
\ee
Taking into account (13) one obtains the relations
\be
P^{(1)} P_j^\pm =P_j^\mp P^{(1)}, \quad P^{(1)}P_3^\pm =P_3^\pm P^{(1)},
\quad T^{(2)} P_a^\pm =P_a^\pm T^{(2)}, \quad j=1,2,
\ee
\be
CP_1^\pm=P_1^\mp C, \qquad CP_2^\pm =P_2^\pm C,
\qquad CP_3^\pm =P_3^\pm C.
\ee

From (14), (16) it follows that

\vspace{-2mm}

\begin{enumerate}
\item[1)]  the system of equations (1), (2) is $T^{(2)}$, $P^{(1)}$, $C$-invariant but $P^{(1)}$,
$C$-non\-in\-va\-ri\-ant,

\vspace{-2mm}

\item[2)] the system of equations (1), (3) is $T^{(2)}$, $C$-invariant but $P^{(1)}$-noninvariant,

\vspace{-2mm}

\item[3)] the system of equations (1), (4) is $T^{(2)}$, $P^{(1)}$-invariant but  $C$-noninvariant.
\vspace{-2mm}
\end{enumerate}

To obtain these result we have used only the relations (13) which are valid for massless
particles of arbitrary spin. The above discussion is followed by tins conclusion:
{\it if the particle (and antiparticle) of zero mass is characterized by helicity and by the sign
of energy only (without additional quantum numbers) three and only three types
of two-component Poincar\'e-invariant essentially different (in respect to C, P, T propertis)
equations exist}. It is interesting to note that the hypothesis of Lee and Yang and Landau
on $CP$-parity conservation is not valid for the equations (1), (3); (1), (4).
Moreover the system of equations (1), (3) is $CP^{(1)}T^{(2)}$-
and $CP^{(1)}T^{(1)}$-noninvariant.
\smallskip

\noindent
{\bf Note 1.} Equation (1) with subsidiary conditions
\be
P_2^\varepsilon P_3^{\varepsilon'} \Psi=0, \qquad \varepsilon,\varepsilon'=\pm1,
\ee
\be
P_2^\varepsilon P_3^{\varepsilon'}\Psi=\Psi,
\ee
is equivalent to three- and one-component equations
\be
(\gamma_\mu p_\mu +\bar \varkappa_0 P_2^\varepsilon P_3^{\varepsilon'})
\varphi^{\varepsilon\varepsilon'}=0, \qquad \varphi^{\varepsilon\varepsilon'}=
\frac 12 (1-P_2^\varepsilon P_3^{\varepsilon'})\Psi, \qquad \mu =0,1,2,3,
\ee
\be
(\gamma_\mu P_\mu +\bar \varkappa_1 P_2^\varepsilon P_3^{-\varepsilon'} +
\bar \varkappa_2 P_2^{-\varepsilon} P_3^{\varepsilon'}+
\varkappa_3 P_2^{-\varepsilon} P_3^{-\varepsilon'})\bar \varphi^{\varepsilon \varepsilon'}=0,
\qquad \bar \varphi^{\varepsilon\varepsilon'}=P_2^\varepsilon P_3^{\varepsilon'}\Psi,
\ee
respectively, where $\bar \varkappa_\mu$ are arbitary constants. It is not difficult to calculate
that there are fifteen equa\-tions (2)--(4), (16), (17) exhausting all possible nonequivalent
Poincar\'e-invariant subsidiary conditions which can be imposed on $\{\Psi\}$.

\smallskip

\noindent
{\bf Note 2.} If a zero-mass particle is characterized by two (but not by one) quantum num\-bers,
there exist more than three types of nonequivalent two-component equations.
Theoretically such a possibility exists due to commutativity of Dirac's Hamiltonian for a
particle of spin $\frac 12$ with $SO_4\sim SU_2\otimes SU_2$ algebra. It means that besides
the mass two conserved quantum numbers $s$ and $\tau$ exist. For the zero-mass case
the eigenvalues of helicity-type operators
\be
\ba{l}
\ds \hat \Lambda_1 =\frac{S_a p_a}{p}, \qquad \hat \Lambda_2 =\frac{\tau_a p_a}{p},
\vspace{2mm}\\
\ds S_a =\frac 12 \left(\frac 12 \varepsilon_{abc} S_{bc} +S_{4a}\right), \qquad
\tau_a =\frac 12 \left(\frac 12 \varepsilon_{abc} S_{bc} -S_{4a}\right)
\ea
\ee
are conserved. If the massless particle is characterized by cigenvalues of operators (20),
the number of theoretically possible equations increases. This follows from the fact that
the two-dimensional irreducible representation of the group $P(1,3)$ for $m\not= 0$
is reduced in the case $m=0$ to the following direct sum of one-dimensional irreducible
representations:
\be
\ba{l}
\ds D^\pm\left(0,\frac 12\right) \to D^\pm \left(0,+\frac 12\right)\oplus D^\pm \left(0,-\frac 12\right),
\vspace{2mm}\\
\ds D^\pm\left(\frac 12,0\right) \to D^\pm \left(+\frac 12,0\right)\oplus D^\pm \left(-\frac 12,0\right).
\ea
\ee
We shall not analyze all possible equations in this case (it is difficult to do this using the
results of paper~[8]) because it is not clear from the physical point of view how one can
distinguish, say, the representations $D^\pm\left(0,-\frac 12\right)$ and
$D^\pm \left(-\frac12, 0\right)$.

\smallskip

{\bf 2.} Let us now show that four- and two-component equations obtained in~[4,~5]
are isometrically equivalent to the Dirac equation (1) and to the Weyl equation.

Consider the four-component equation of the type [4]
\be
i\frac{\p \Phi(t,\pbf{x})}{\p t}={\mathcal H}_\Phi \Phi(t,\pbf{x})=(\alpha_a p_a +
\Lambda)\Phi(t,\pbf{x}), \qquad \alpha_a =\gamma_a \gamma_a,
\ee
where $\Lambda$ is an operator satisfying the condition
\be
\alpha_a p_a \Lambda=-\Lambda \alpha_a p_a, \qquad \Lambda^2=0.
\ee
Equation (22) can be obtained from (1) with the help of the isometric transformation
\be
\Psi\to \Phi =V_1\Psi, \qquad {\mathcal H}\to {\mathcal H}_\Phi=V_1{\mathcal H} V_1^{-1},
\ee
where
\be
V_1=1-\frac 12 \frac{\alpha_a p_a}{E^2}\Lambda, \qquad
V_1^{-1}=1+\frac 12 \frac{\alpha_a p_a}{E^2}\Lambda.
\ee
The Hamiltonian ${\mathcal H}_\Phi$ is Hermitian in respect of the following scalar product:
\be
(\Phi_1,\Phi_2)=\int d^3\pbf{x}\; \Phi_1^\dag (t, \pbf{x})(V_1^{-1})^\dag V_1^{-1} \Phi_2(t,\pbf{x}).
\ee
To draw the correct conclusion about the $C$, $P$, $T$ propertios equation (22) it is necessary
to write the algebra (5) in the $\Phi$-representation. We shall not do this here. We shall remark
only that due to the invariance of equation~(1) under $P^{(1)}$, $T^{(2)}$,
$C$ transformations equation (22) is invariant with respect to the transformations
\be
P_\Phi^{(1)} =V_1 P^{(1)} V_1^{-1}, \qquad C_\Phi=V_1C V_1^{-1},
\qquad T_\Phi^{(2)}=V_1 T^{(2)} V_1^{-1}.
\ee

One can show in an analogous manner that the two-component equation of the type~[5]
\be
\ba{l}
\ds i\frac{\p \chi(t,\pbf{x})}{\p t} =(\sigma_ap_a+B)\chi(t,\pbf{x}),
\vspace{2mm}\\
\ds B\sigma_a p_a=-\sigma_a p_a B, \qquad B^2=0,
\ea
\ee
can be obtained from the Weyl equation with the help of the operator
\be
V_2 =1-\frac 12 \frac{\sigma_ap_a}{E^2}B, \qquad
V_2^{-1} =1+\frac 12 \frac{\sigma_ap_a}{E^2}B.
\ee

\medskip
\begin{enumerate}

\footnotesize

\item  Fushchych W.I., {\it Nucl. Phys. B}, 1970, {\bf 21}, 321, {\tt
quant-ph/0206077};\\ Fushchych W.I., {\it Theor. Math. Phys.},
1971, {\bf 9}, 91 (in Russian).

\item Fushchych W.I.,  Grishchenko A.L., {\it Lett. Nuovo Cimento}, 1970, {\bf 4}, 927, {\tt quant-ph/0206078}.

\item Simon M.T., {\it Lett. Nuovo Cimento}, 1971, {\bf 2}, 616.

\item Santhanam T.S., Tekumalla A.R., {\it Lett. Nuovo Cimento}, 1972, {\bf 3}, 190.

\item  Tekumalla A.R., Santhanam T.S., {\it Lett. Nuovo Cimenro}, 1973, {\bf 6}, 99.

\item Seetharaman T.S., Simon M.T., Mathews P.M., {\it Nuovo Cimetro A}, 1972, {\bf 12}, 788.

\item  Fushchych W.I., Grishchenko A.L., Nikitin A.G., {\it Theor. Math. Phys.}, 1971,
{\bf 8}, 192 (in Rus\-sian).

\item Fushchych W.I., {\it Lett. Nuovo Cimento},  1973, {\bf 6}, 133, {\tt quant-ph/0206105}.
\end{enumerate}
\end{document}